\documentclass{article}
\usepackage[margin=1in]{geometry}

\usepackage{amsmath,amsfonts,amssymb}
\usepackage{graphicx}
\usepackage[colorlinks=true, allcolors=blue]{hyperref}
\usepackage{caption}
\usepackage{subcaption}
\usepackage{lipsum}
\usepackage[version=4]{mhchem}
\usepackage{gensymb}
\usepackage{authblk}
 
\begin{document} 

\captionsetup[subfigure]{justification=justified,singlelinecheck=false}
\newcommand{\varsizetwo}{20pt}
\newcommand{\varsizeone}{1.45in}

\title{Hybrid integrated laser at visible wavelengths using aluminum nitride photonic integrated circuit}

\author[a,b]{Nikolay Videnov\footnote{nikolay.videnov@uwaterloo.ca}}
\author[a,c]{Matthew L. Day}
\author[a,b,c]{Michal Bajcsy}

\renewcommand{\Affilfont}{\small \itshape}

\affil[a]{Institute for Quantum Computing, 200 University Ave W, Waterloo, ON N2L 3G1}
\affil[b]{Department of Electrical and Computer Engineering, 200 University Ave W, Waterloo, ON N2L 3G1}
\affil[c]{Department of Physics and Astronomy, 200 University Ave W, Waterloo, ON N2L 3G1}

\maketitle

\renewenvironment{abstract}
 {\small 
  \list{}{%
    \setlength{\leftmargin}{1in}% <---------- CHANGE HERE
    \setlength{\rightmargin}{\leftmargin}%
  }%
  \item\relax}
 {\endlist}

\begin{abstract} 
\textbf{
We show the first demonstration of a hybrid external cavity diode laser (ECDL) using aluminum nitride (AlN) as the wave-guiding material. Two devices are presented, a near-infrared (NIR) laser using a 850~nm diode and a red laser using a 650~nm diode. The NIR laser has $\approx$1~mW on chip power, 6~nm of spectral coverage, instantaneous linewidth of 720$\pm$80~kHz, and 12~dB side mode suppression ratio (SMSR). The red laser has 15~dB SMSR. }
\end{abstract}

\section{INTRODUCTION} \label{sec:intro}

Aluminum nitride (AlN) is a high index (2.2) and large band-gap (6~eV) material. Both properties are necessary for tightly confining and low loss waveguides that can operate at wavelengths used in atom-cooling and ion-trapping experiments, which range from ultra-violet (UV) to infra-red (IR). AlN also has an electro-optical coefficient, optical non-linearities ($\chi^{(2)}$ and $\chi^{(3)}$), and is piezoelectric. These features have been harnessed into many applications, well documented in reviews by N. Li \emph{et al.}\cite{Li2021} and Liu \emph{et al.}\cite{Liu2023}.

A device not featured in these reviews is a fully on-chip light source. The benefits of miniaturization and scaleability for photonics are somewhat diminished if the light source is off-chip. As photonic circuits become more complicated additional light sources are needed and can become a limiting factor in design. One method of producing narrow linewidth laser light, on-chip, is to use a hybrid external cavity diode laser (ECDL).

The typical construction for an ECDL is a laser diode with a bulk-opitcs grating providing feedback to the diode which determines the lasing characteristics. A hybrid ECDL miniaturizes this idea by edge coupling a laser diode into a planar photonic integrated circuit (PIC). Conceptually shown in Figure~\ref{fig:concept}. The function of the grating is performed by the PIC chip. Hybrid ECDLs are known for similar spectral qualities to bulk ECDLs with the added benefits of a PIC platform. Recent exemplars include publications by Van Rees \emph{et al.}\cite{VanRees2023} and M. Li \emph{et al.}\cite{Li2022}. Both of these publications operate in the telecommunications wavelength range partially due to the material choice. Silicon nitride (\ce{SiN_x}) and lithium niobate (\ce{LiNbO3}) are the two popular material platforms for hybrid ECDLs. However, both suffer from a relatively small bandgap (4~eV), and high power circuits are not possible for the whole visible-to-UV spectrum.

Alternative material platforms are required to address this. Franken \emph{et al.}\cite{Franken2023} use alumina (\ce{Al2O3}); another high index (1.8) and large bandgap (7~eV) material. Alumina is excellent for UV waveguides but thanks to the presence of optical non-linearities AlN is more versatile with many active and non-linear devices having been demonstrated. This paper describes the first demonstration of a hybrid ECDL where the wave-guiding material is aluminum nitride.

\section{FABRICATION} \label{sec:design}  

\begin{figure}
    \centering
    \begin{subfigure}[t]{.55\textwidth}
        \caption{}
        \label{fig:crossSection}
        \includegraphics[width=\textwidth]{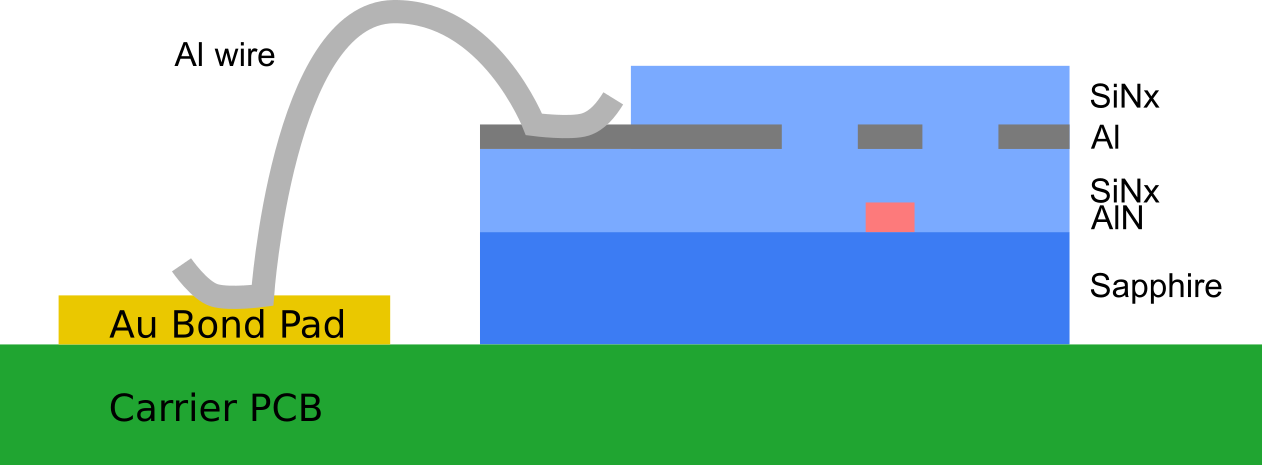}
    \end{subfigure}~
    \begin{subfigure}[t]{.35\textwidth}
        \caption{}
        \label{fig:SEMsidewall}
        \includegraphics[width=\textwidth]{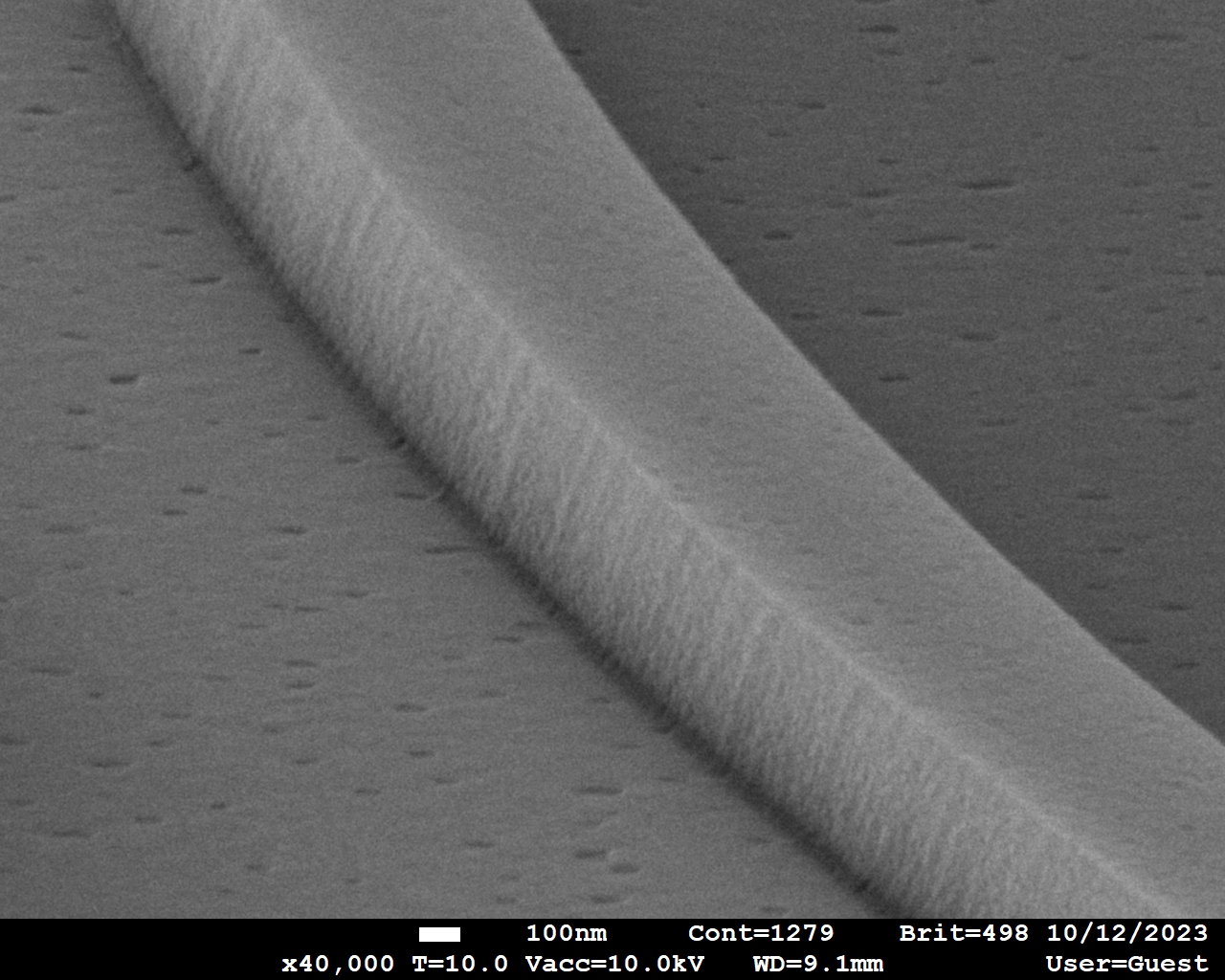}
    \end{subfigure}\\
    \begin{subfigure}[t]{.4\textwidth}
        \caption{}
        \label{fig:TETM}
        \includegraphics[width=\textwidth]{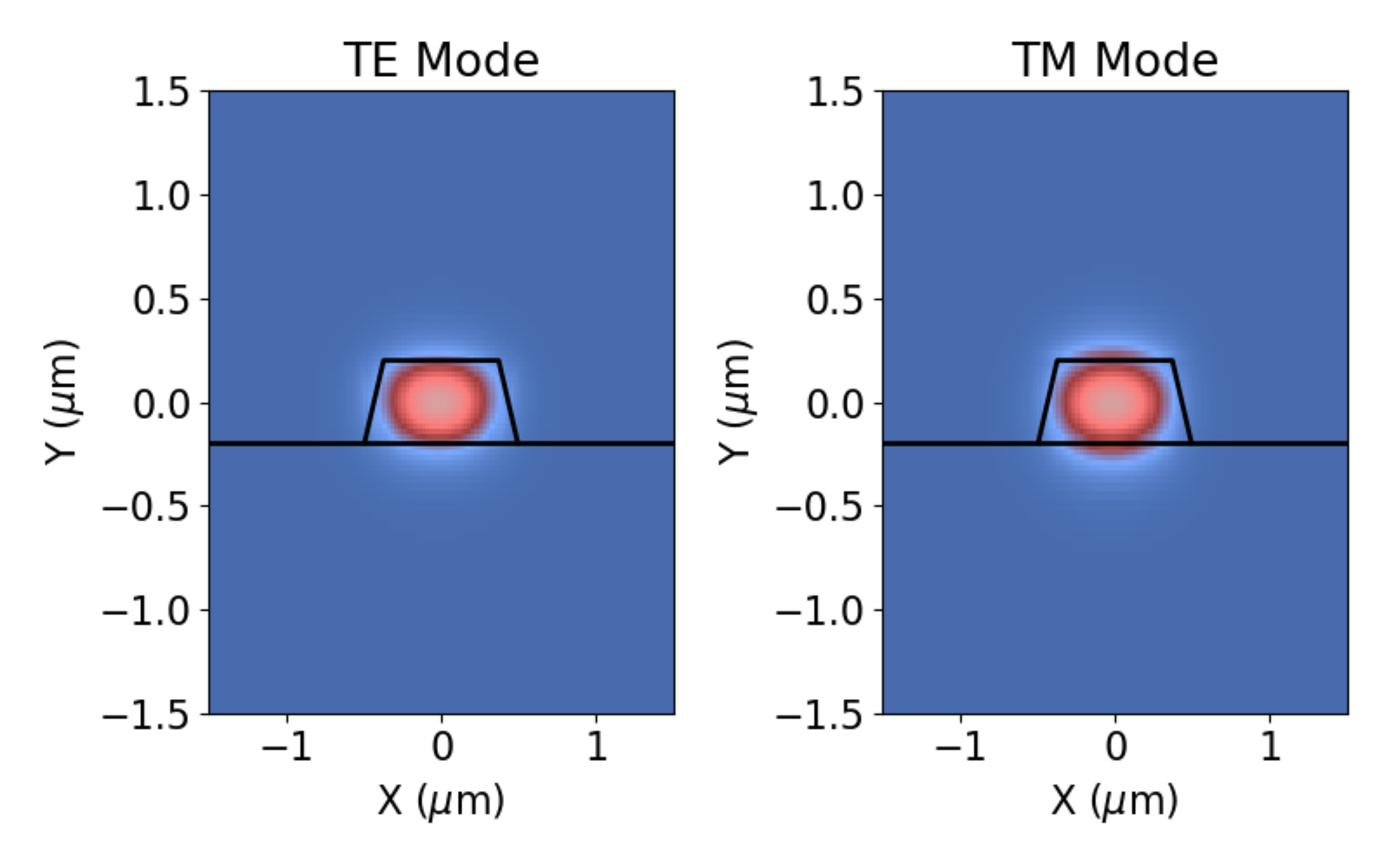}
    \end{subfigure}~
    \begin{subfigure}[t]{.55\textwidth}
        \caption{}
        \label{fig:experimentalSetup}
        \includegraphics[width=\textwidth]{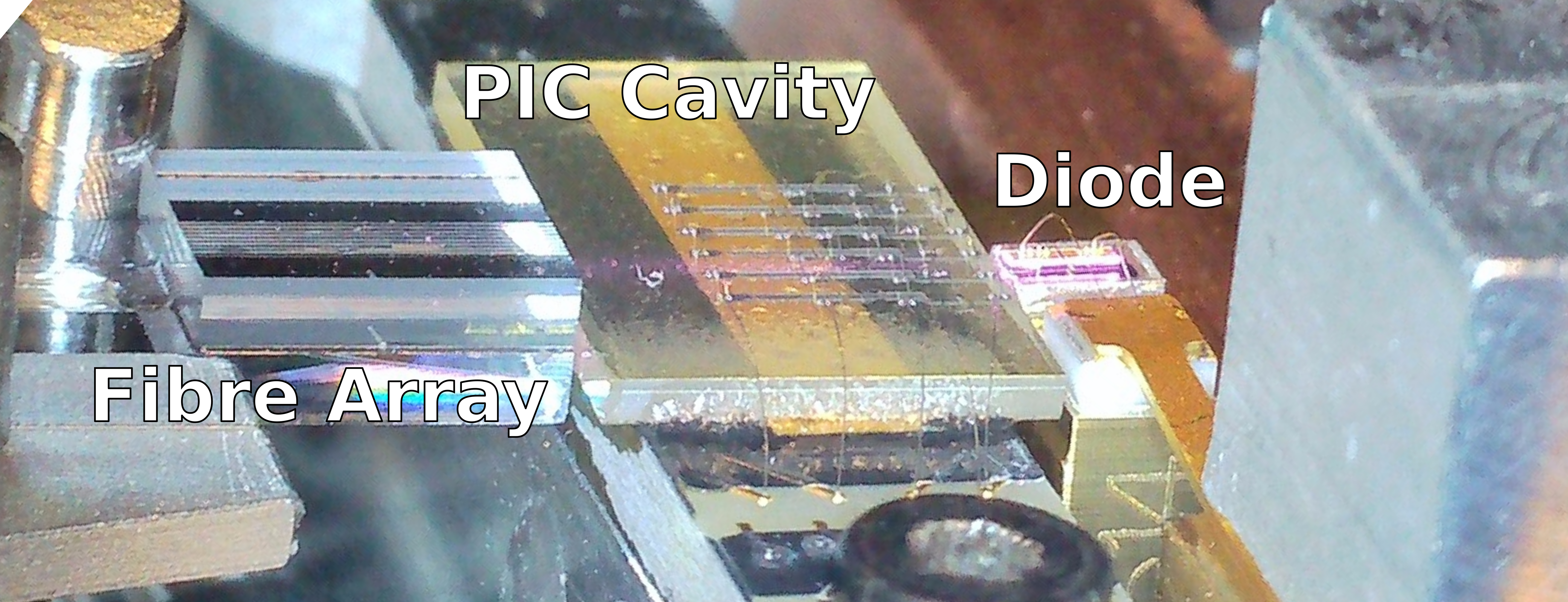}
    \end{subfigure} 
    \caption{(a) Cross section of a fabricated device showing material stack. Electrodes can be clad if high voltages are required. (b) SEM image of waveguide showing sidewall texture. Image taken at 10$\degree$. No surface preparation was done prior to imaging. (c) Lumerical MODE simulation of TE and TM modes. The pictured waveguides and modes are for the NIR laser. (d) Testing setup for presented devices. On the left is a fibre array with 6 SM and 2 MM fibres held at a 127~$\mu$m pitch. In the middle is the PIC chip. On the right is a c-mount laser diode from Sacher Lasertechnik. The PIC chip is edge coupled on both sides and mounted to a carrier PCB.}
    \label{fig:dummy2}
\end{figure}

The waveguide dimensions were nominally 400~nm x 750~nm for the NIR laser and 200~nm x 750~nm for the red laser. This was found to produce suitably low loss waveguides while maintaining tight bending radius ($>$60$~\mu$m) and single mode operation. Simulation using Lumerical MODE validated the bending radius and single mode status. See Figure~\ref{fig:TETM} for calculated TE and TM mode profiles.

The feedback cavity was designed to have a free spectral range (FSR) around 7.5~nm. Figure~\ref{fig:filter} shows the as-designed filter which was simulated using Lumerical INTERCONNECT. Each micro-ring resonator was fabricated to be considerably larger than the minimum bend radius for ease of construction and tuning. Improved performance can be found by reducing ring length. 

\renewcommand{\varsizetwo}{20pt}
\renewcommand{\varsizeone}{1.45in}
\begin{figure}
    \centering
    \begin{subfigure}[t]{.4\textwidth}
        \caption{}
        \label{fig:concept}
        \includegraphics[width=\textwidth]{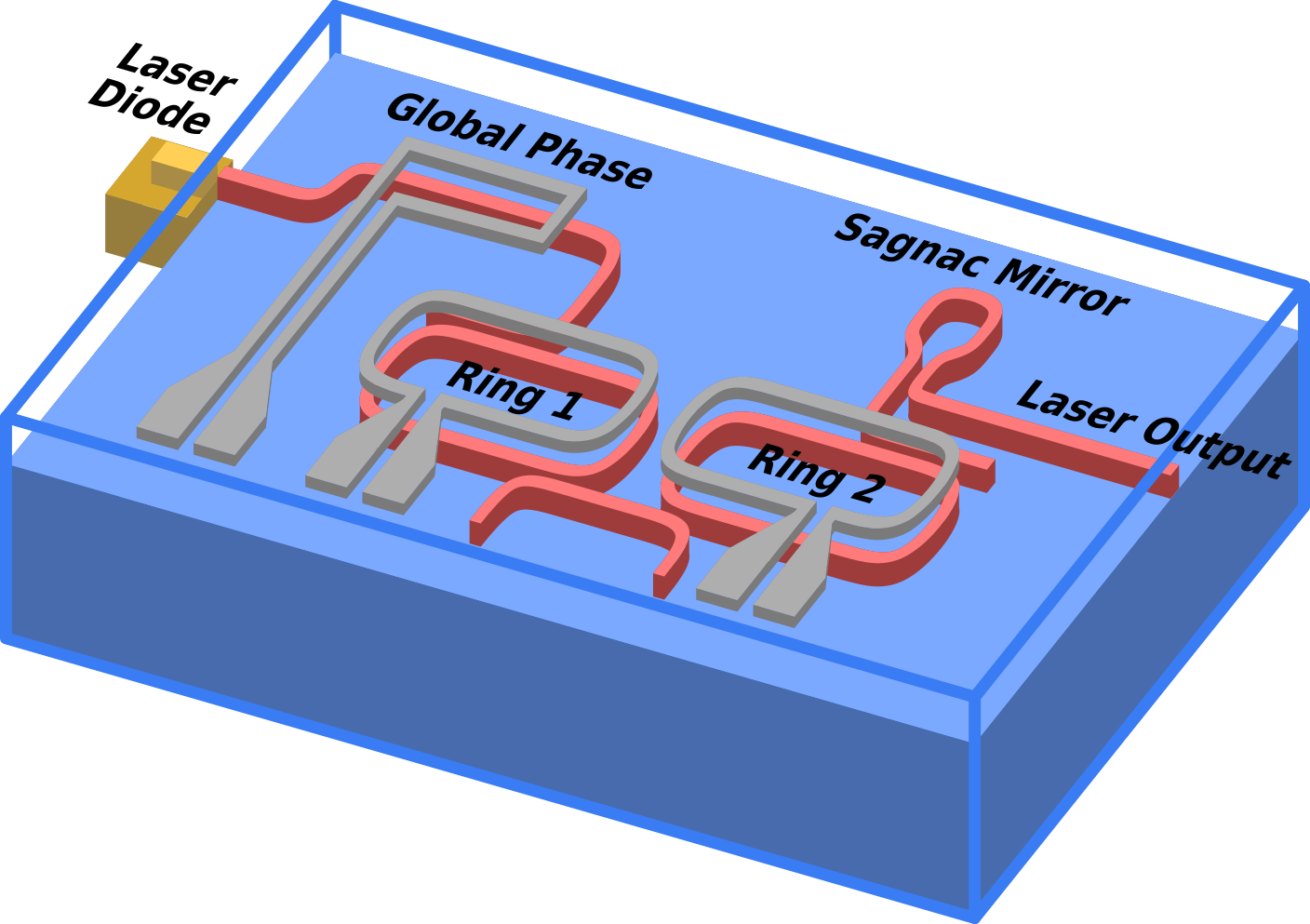}
    \end{subfigure}\hspace{\varsizetwo}
    \begin{subfigure}[t]{.4\textwidth}
        \caption{}
        \label{fig:filter}
        \includegraphics[width=\textwidth]{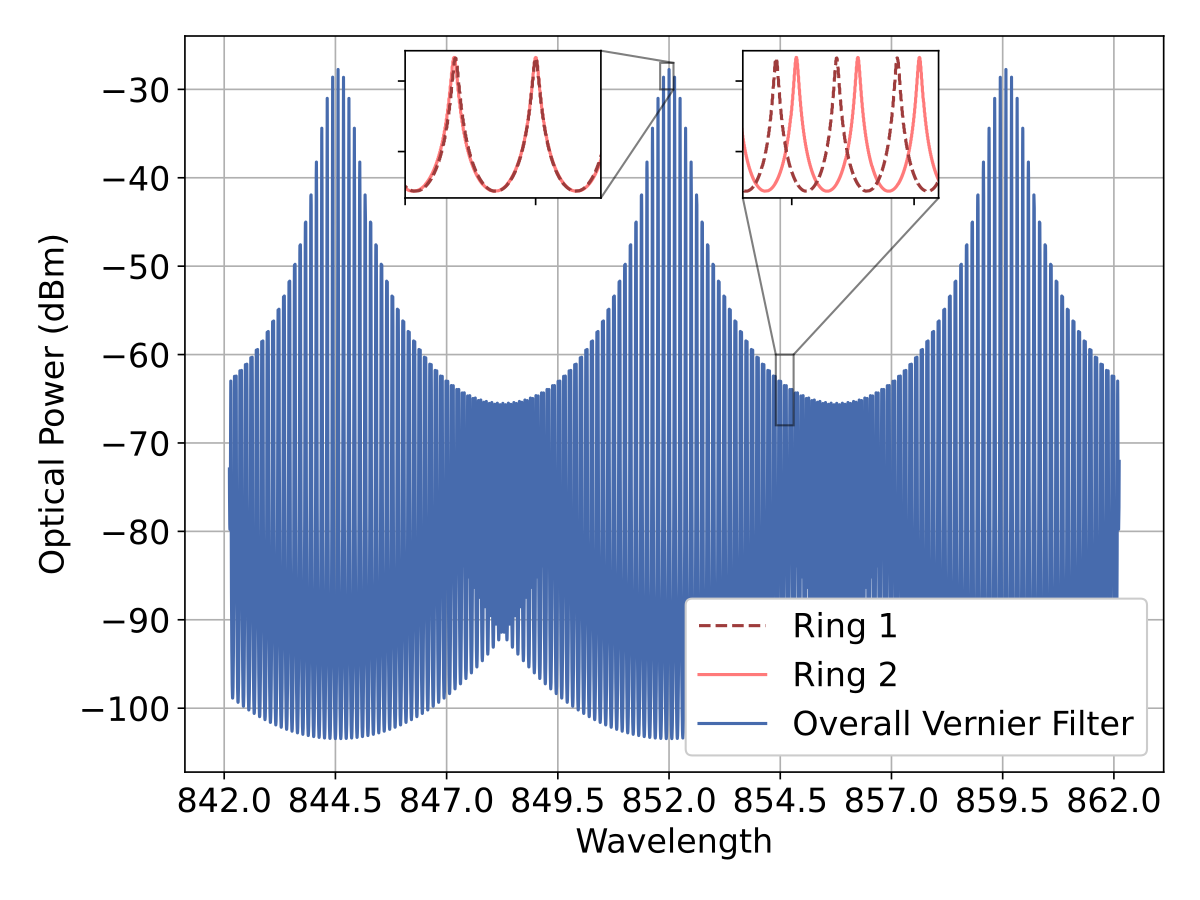}
    \end{subfigure}\\
    \caption{(a) Concept diagram. Shown are the main aspects of a hybrid ECDL, and configuration for the fabricated devices. In gold is the gain medium (laser diode) which is edge coupled into the PIC. There are three control electrodes in the filter, from left to right: global phase, ring 1, and ring 2. The final element is the Sagnac mirror which partially reflects light back to the laser diode. (b) Simulated vernier filter. Formed by two cascaded micro-ring resonators with a small length mismatch. See inserts showing how the length mismatch creates the filter shape. Each ring has modest FSR; the cascaded rings achieve a 7.5~nm FSR.}
    \label{fig:dummy1}
\end{figure}

Standard lithographic techniques were used to fabricate the waveguides. Commercial wafers of thin film AlN on sapphire were purchased from Kyma Technologies. The waveguides were patterned through electron beam lithography and transferred into the AlN layer through reactive ion etching. A low index \ce{SiN_x} cladding was deposited using plasma enhanced chemical vapour deposition. \ce{SiO2} is also available, however \ce{SiN_x} is easier to work with while the devices are far from the absorption limit for \ce{SiN_x}. Electrodes were patterned on top of the cladding through UV photo-lithography and lift-off. Optical quality facets were made by dicing and polishing. The PIC chip was mounted on a carrier printed circuit board (PCB) which provides a thermal sink. Electrical connections were made by wire bonding from the surface of the PIC chip to the PCB. The cross section of a finished device is shown conceptually in Figure~\ref{fig:crossSection}.

Propagation loss is the figure of merit for waveguides. This limits circuit length and cavity quality factor. Consequently setting a limit on laser linewidth\cite{Boller2019}. The main source of propagation losses is sidewall roughness. The fabrication recipe is evaluated by the produced roughness and propagation loss.

Figure~\ref{fig:SEMsidewall} shows a SEM image of an AlN waveguide. There is visible texture on the sidewall. The magnitude of the roughness is small leading to propagation losses equal to 3.9$\pm$0.8~dB/cm in the NIR laser presented here. There is variation in loss from chip-to-chip with the best measured loss being 2.0$\pm$0.3~dB/cm at 852~nm. For the red laser the loss was not characterized, though typical propagation losses for this recipe at 650~nm are 6~dB/cm.

\renewcommand{\varsizetwo}{20pt}
\renewcommand{\varsizeone}{1.45in}

\newpage
\section{RESULTS} \label{sec:results}  

\renewcommand{\varsizeone}{.4}
\renewcommand{\varsizetwo}{20pt}
\begin{figure}
    \centering
    \begin{subfigure}[t]{\varsizeone\textwidth}
        \caption{}
        \label{fig:singleModelasing}
        \includegraphics[width=\textwidth]{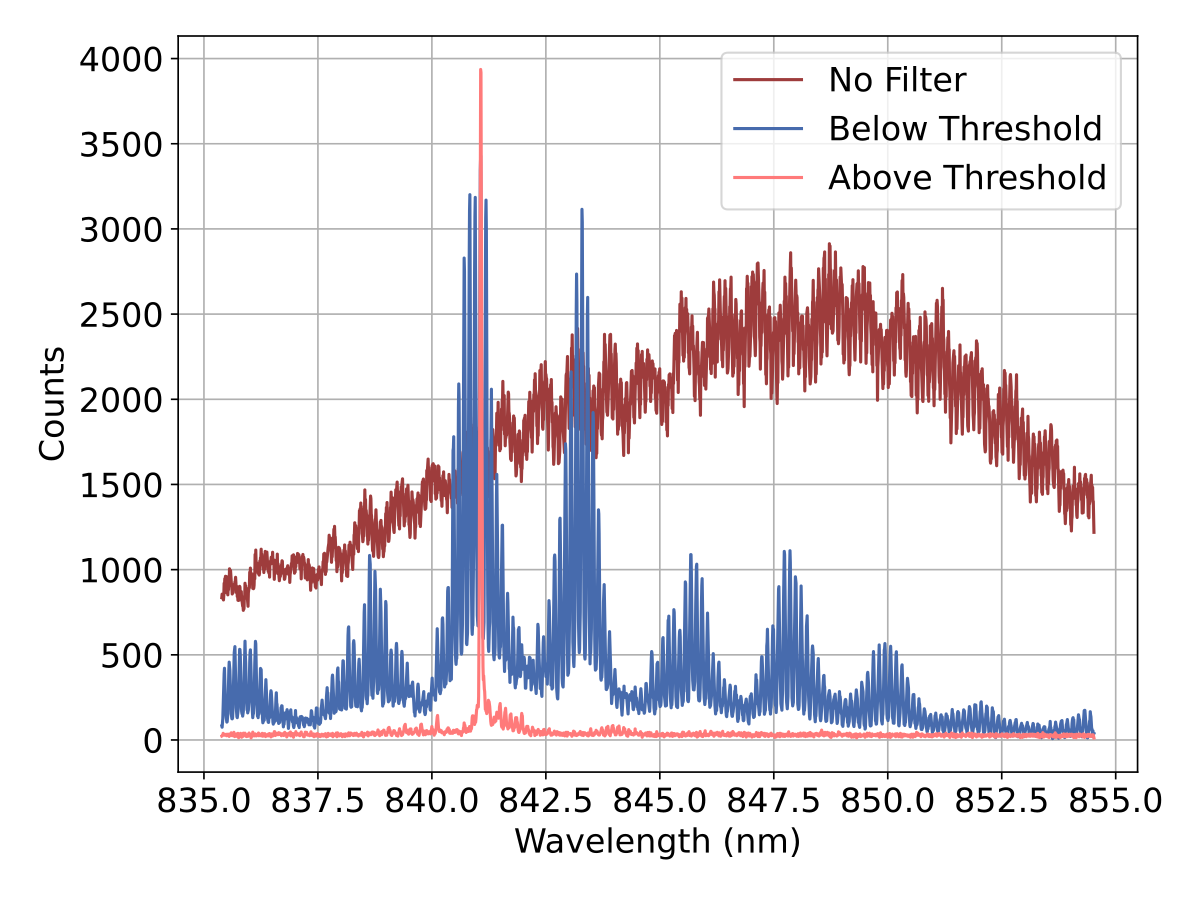}
    \end{subfigure}
    \hspace{\varsizetwo}
    \begin{subfigure}[t]{\varsizeone\textwidth}
        \caption{}
        \label{fig:redLaser}
        \includegraphics[width=\textwidth]{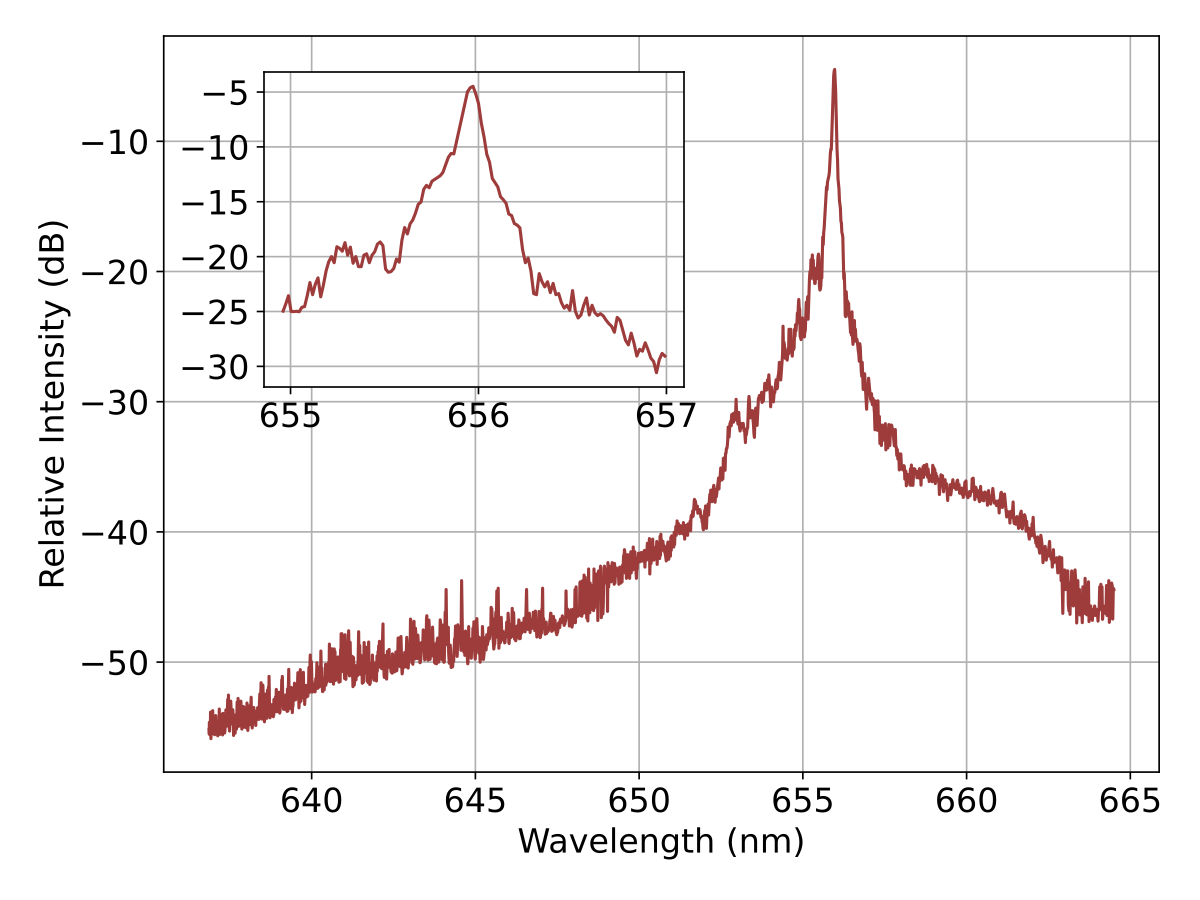}
    \end{subfigure}\\
    \begin{subfigure}[t]{\varsizeone\textwidth}
        \caption{}
        \label{fig:tuning}
        \includegraphics[width=\textwidth]{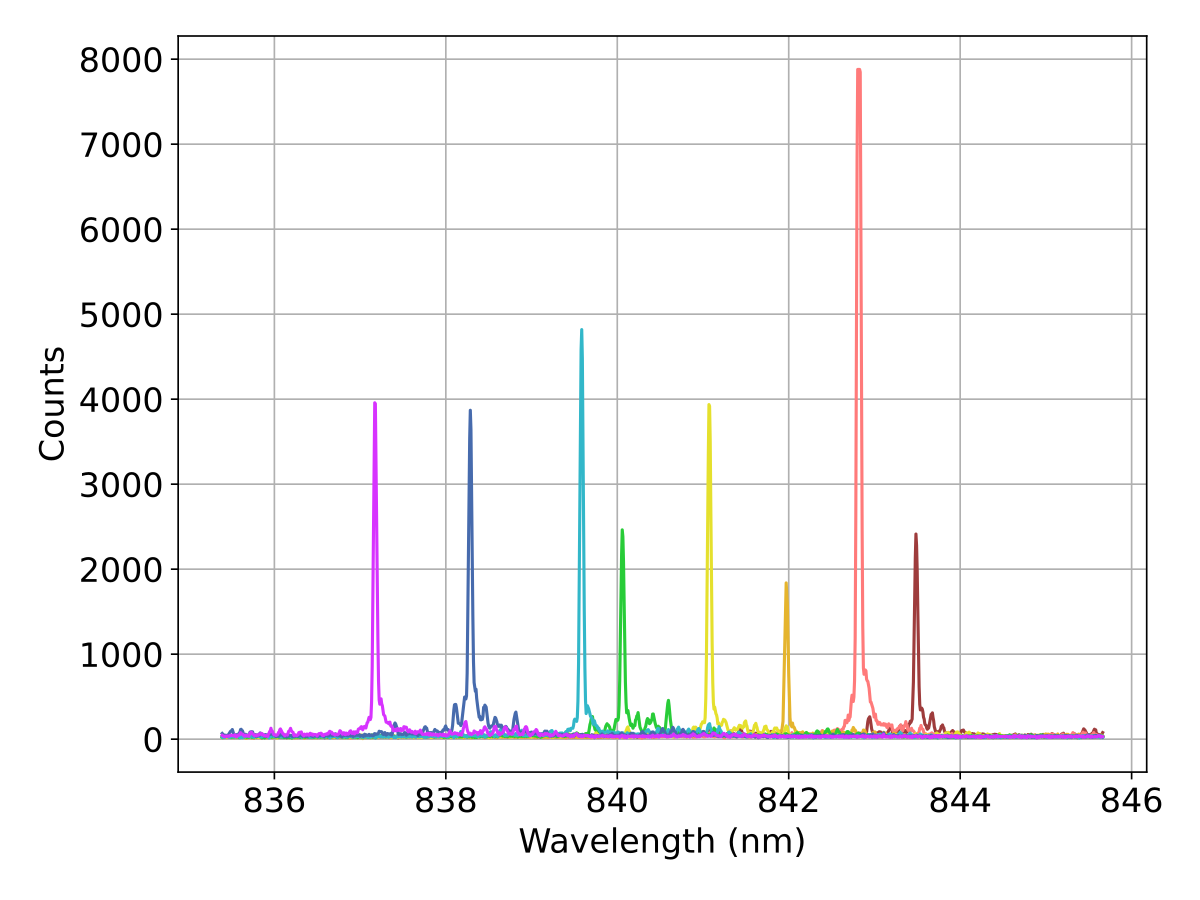}
    \end{subfigure}
    \hspace{\varsizetwo}
    \begin{subfigure}[t]{\varsizeone\textwidth}
        \caption{}
        \label{fig:linewidth}
        \includegraphics[width=\textwidth]{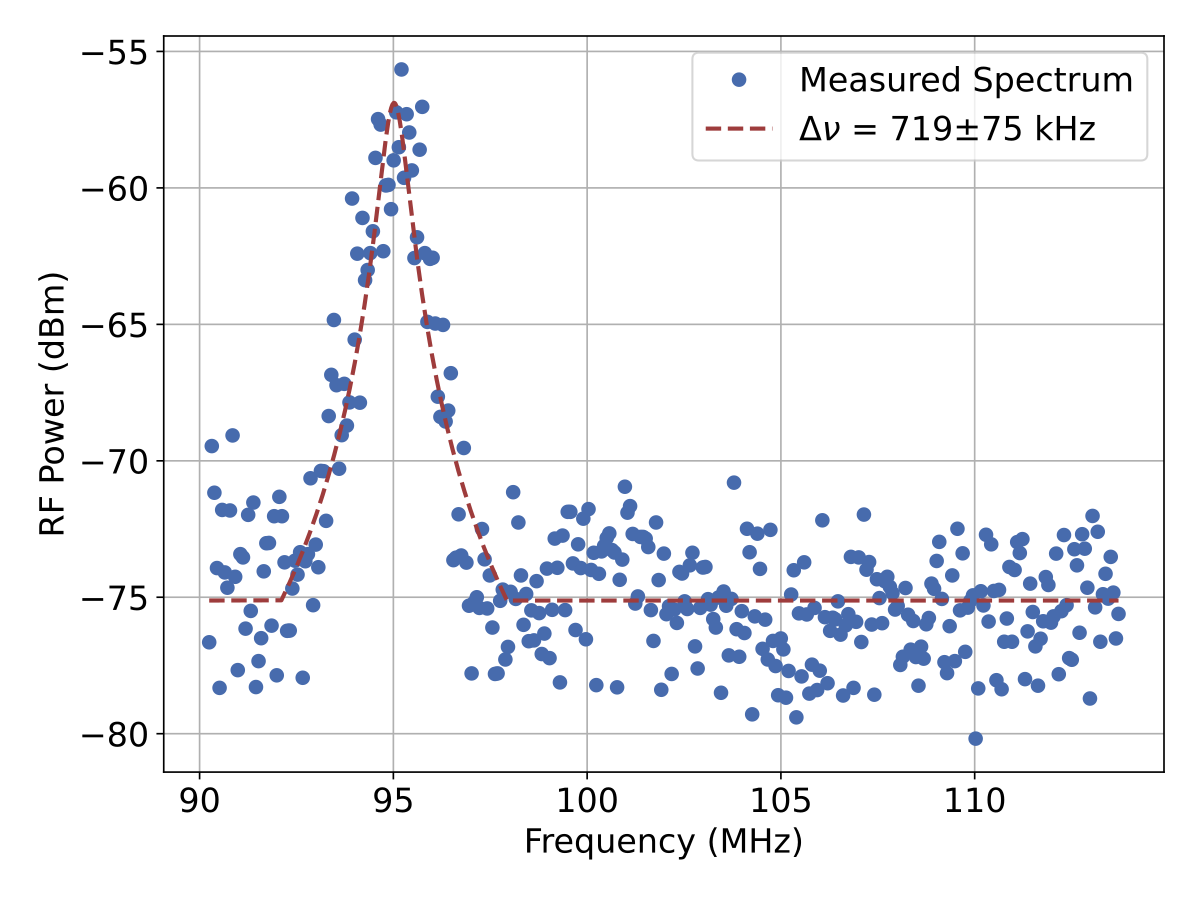}
    \end{subfigure}
    \caption{(a) Spectrometer reading of the NIR laser. In red is the laser diode without feedback from the cavity. In blue is the NIR laser below lasing threshold; the competing modes of the vernier filter are visible. In pink is the NIR laser above lasing threshold. (b) High dynamic range spectrometer image of the red laser. The competing side-modes are suppressed by 15~dB. Spectrometer resolution is 100~MHz; not suitable for linewidth measurement. (c) Tuning of lasing wavelength by 6~nm. Achieved by changing the ring 1 index with a thermal tuner. Maximum applied thermal power was 25~mW. (d) Measured using a Rigol DSA815 spectrum analyzer with 30~kHz RWB and VWB. The noise floor for this measurement bandwidth is $-$75~dBm. Data fit with Lorentzian function.}
    \label{fig:dummy3}
\end{figure}

Two devices were fabricated lasing in the NIR and in the red. The NIR laser used an 850~nm laser diode while the red laser used a 650~nm laser diode. The diodes were anti-reflection coated on the facet side and high-reflection coated on the opposing side. The NIR laser was fabricated with control electrodes and was tested thoroughly. The red laser is a proof of principle for operation at shorter wavelengths in the visible spectrum.

The optical spectrum of both devices was recorded using a Moglabs MWM spectrometer, shown in Figure~\ref{fig:singleModelasing}-Figure~\ref{fig:tuning}. The NIR laser spectrum is plotted without any feedback (only the diode), with cavity feedback but below threshold, and above lasing threshold. There is a distinct change from multi-mode to single-mode operation. For both devices a high dynamic range (HDR) spectrum was taken, Figure~\ref{fig:redLaser} shows the red laser HDR spectrum. The side mode suppression ratio (SMSR) was measured to be 15~dB for the red laser, and 12~dB for the NIR laser.

Three micro-heaters were fabricated on the NIR laser. One heater for overall phase, and one on each micro-ring resonator as depicted in Figure~\ref{fig:concept}. Heating only one ring caused the filter to skip ``comb teeth'' and produced large changes in lasing wavelength. In Figure~\ref{fig:tuning} the lasing wavelength was pulled by 6~nm by applying 25~mW of heat to the first ring.

The instantaneous laser linewidth of the NIR laser was measured by beating the fabricated laser against a M SQUARED ultra narrow linewidth CW Ti:Sapphire laser. The radio frequency beat-note is shown in Figure~\ref{fig:linewidth}. The linewidth of the Ti:Sapphire laser has not been measured, but normal operation should be much narrower than the fabricated device. The NIR laser had an instantaneous linewidth of 720$\pm$80~kHz. This measurement was done with around 1~mW of on-chip power.

\section{DISCUSSION}\label{sec:discussion}

Neither of the two lasers, from the beat-note measurement, were actively stabilized. There was significant slow drift between the two lasers which is not accounted for in the pictured spectrum due to the short single-shot measurement time of the spectrum analyzer.

The NIR laser tuning was solely through micro-heaters. This was a significant limitation for tuning. The target lasing wavelength was 852~nm, however due to manufacturing inconsistencies the cold cavity resonance was around 840~nm. Applying a current to either ring tuner can only pull the filter to shorter wavelengths. This could be mitigated by designing for nominal operation at a longer wavelength. The micro-heaters take $\approx$40s to stabilize; quite slow. Maintaining a set-point requires constant current flow, and constant power consumption.

An advantage of AlN is the non-zero electro-optical (EO) coefficient. Tuning could instead be done through EO phase shifters which operate at GHz timescales\cite{Li2022}. The power consumption of maintaining a constant voltage is negligible. The change in refractive index is also bi-directional. Unlike a thermal tuner which can only heat up, an EO tuner can have positive or negative voltage. The EO coefficient for these thin films was measured to be 0.625$\pm$0.003~pm/V and 1.079$\pm$0.003~pm/V for $r_{13}$ and $r_{33}$. The drawback to EO phase shifters in AlN is the modest EO coefficient producing limited tuning range.

The in-fibre power was around 60~$\mu$W. Coupling losses between chip and fibre are typically $-$15~dB, this device had measured losses of $-$10~dB on the opposite facet to the laser output. From the coupling loss the estimated on-chip power is between 0.6~mW and 1.9~mW. The edge couplers in use for this design are simple up-tapers. This limits what control mechanisms can be implemented for frequency stabilization, and limits the characterization schemes. Straightforward design changes can improve coupling loss. Either in the same waveguide layer\cite{Liu2020} or by adding a dedicated layer\cite{Lin2021}.

Comparing the simulated and measured (Figure~\ref{fig:filter} and Figure~\ref{fig:singleModelasing}) vernier filters there is a discrepancy. The relative amplitude between peaks is not expected to match simulation. The light source is the laser diode which has a non-uniform intensity. The measured FSR is around 2.5~nm, while the simulation predicts 7.5~nm. This discrepancy can be explained by a non-uniform refractive index across the wafer. Future fabrication runs will have a larger parameter space for post-selecting.

\section{CONCLUSION} \label{sec:conclusion}  

A hybrid ECDL was fabricated using an AlN material platform for the first time. Two devices are shown in this paper using 850~nm and 650~nm laser diodes. The NIR laser was thoroughly characterized showing 6~nm of spectral coverage, 720$\pm$80~kHz linewidth, 0.6-1.9~mW on-chip power, and 12~dB SMSR. The red laser showed promise of future extension to shorter operating wavelength.

\section*{ACKNOWLEDGMENTS}       

The University of Waterloo's QNFCF facility was used for this work. This infrastructure and the presented work would not be possible without the significant contributions of CFREF-TQT, CFI, ISED, the Ontario Ministry of Research \& Innovation and Mike \& Ophelia Lazaridis. Their support is gratefully acknowledged.

% References
%\printbibliography
\newpage
\bibliographystyle{ieeetr}
\bibliography{refer}

\end{document}